\documentclass[a4paper,12pt]{article}
\title{\textbf{Anomalies In The $\beta$ decay Processes And The Pulse Strong-Current
  Discharges As Consequence Of Electron Gravitational Emission}}
                                
\author{S.I. Fisenko \\
           Kazakh State University, Physics Department \\
          Al Farabi str., 71, 480078 Almaty, Kazakhstan \\
                 E-mail:  altair\_alm@hotmail.com}

\date{}

\begin{document}

\maketitle

\begin{abstract}

Parity nonconservation in the $\beta$ decay processes is considered  as
fundamental  property of weak interactions.  Nevertheless,  this
property  can  be  treated as anomaly,  because  in  fundamental
interactions of the rest types parity is conserved. Analogously,
anomaly in the short-duration strong-current pulse discharges is
well   known.  The  essence  of  this  phenomenon  consists   in
generation  of  local high-temperature plasma formations  (LHTF)
with  the  typical  values  of  its  thermodynamical  parameters
exceeding  those related to the central section of a  discharge.
In this paper, an attempt is undertaken to treat these anomalies
as  manifestation  of  fundamental properties  of  gravitational
emission. Some consequences of this assumption can be tested  in
the $\beta$ decay experiments as well as in the experiments with short-
duration $z$-pinch-type pulse discharges.
\end{abstract}

\newpage
\section{Quantum-Level Gravitational Interaction, Limiting
Transition to GRT}

Two  points are of importance for the model considered.  (1)  In
the  Einstein  field  equations $\kappa$   is  a  constant  that  relates
geometric  properties of the space-time to the  distribution  of
physical  matter,  and origin of the equations isn't  associated
with  numerical restriction imposed on the constant $\kappa$  .  However,
the  correspondence  principle  (requirement  of  correspondence
between  the  Relativistic Theory of Gravity and  the  Newtonian
Classic  Theory of Gravity) leads to small value of the constant
$\kappa =8\pi G/c^4$ ,  where $G$ and $c$ are, respectively, the
 Newtonian gravitational
constant and the velocity of light. The correspondence principle
follows from the primary concept of the Einstein GTR treated the
latter as relativistic generalisation of the Newtonian Theory of
Gravity. (2) Equations which incorporate the $\Lambda$   term are the  most
general ones in the Relativistic Theory of Gravity. The limiting
transition to weak fields leads to the equation:

\begin{eqnarray*}
\Delta \Phi =-4\pi \rho G +\Lambda c^2,
\end{eqnarray*}
                                
(here $\Phi$   is  the field scalar potential, $\rho$  is the source density),
rather  than  to  the Poisson equation. This fact,  finally,  is
crucial when neglecting the $\Lambda$ -term, because only in this case GTR
can  be  considered  as generalised Classic Theory  of  Gravity.
Thus, numerical values of the quantities $\kappa = 8\pi G/c^4$ and $\Lambda =0$
  in the GTR aren't
associated  with  origin of equations but  originate  only  from
correspondence between GRT and  the appropriate classic theory.
Beginning  with seventies, it has become clear \cite{1} that  in  the
quantum  region  the  numerical value of the  constant $G$  isn't
compatible with principles of Quantum Mechanics. In a number  of
papers  (\cite{1}, \cite{2}) it was shown that in the quantum  region  the
coupling  constant $K$ is more accessible ($K\approx 10^{40}G$). So  the  problem  of
quantum-level generalisation of relativity equations was reduced
to  matching  the numerical values of gravity constants  in  the
quantum and classic regions.
As  a  development of these results concerning  the  micro-level
approximation  of  the  Einstein field  equations,  a  model  is
proposed under the following assumption:

\emph{The gravitational field within the region of localisation of  an
elementary  particle  having a mass $m_0$ is characterised  by  the
values  of  the gravity constants $K$ and $\Lambda$  that lead to stationary
states  of  the particle in its proper gravitational field,  and
the   particle  stationary  states  are  the  sources   of   the
gravitational field with the Newtonian gravity constant $G$.}

In  the  frame  of the Gravity Theory the most general  approach
takes  twisting into account and treats the gravitational  field
as the gauge one, considered similar to other fundamental fields
\cite{3}. This approach is get rid of a priory grounds as applied  to
geometrical properties of the gravitational field, and it  seems
to  be  reasonable  at a microscopic level. For  the  elementary
source  of a mass $m_0$, the equation set describing its states  in
the  proper  gravitational  field,  according  to  the  accepted
assumption, looks like this:

\begin{eqnarray}
\left\{i\gamma ^{\mu }\left(\nabla _{\mu } + \bar \kappa \Psi \gamma _{\mu }\gamma _5
\Psi \gamma _5 \right) - m_0c/\hbar  \right\}\Psi = 0 
\label{1}
\end{eqnarray}
                                                             
\begin{eqnarray}
R_{\mu \nu }-\frac{1}{2}g_{\mu \nu }R=-\kappa \left\{T_{\mu \nu }(E_n)-\mu g_{\mu \nu }
+\left(g_{\mu \nu }S_{\alpha }S^{\alpha }-S_{\mu }S_{\nu }\right)\right\}
\label{2}
\end{eqnarray}

\begin{eqnarray}
R\left(K,\Lambda ,E_n,r_n\right)=R\left(G,E'_n,r_n\right)
\label{3}
\end{eqnarray}
                                                             
\begin{eqnarray}
\left\{i\gamma ^{\mu }\nabla _{\mu } - m_nc/\hbar  \right\}\Psi ' = 0
\label{4}
\end{eqnarray}

\begin{eqnarray}
R_{\mu \nu }-\frac{1}{2}g_{\mu \nu }R=-\kappa ' T_{\mu \nu }\left(E'_n\right)
\label{5}
\end{eqnarray}
                                                             
The following notations are used throughout the article:$\kappa =
8\pi K/c^4$, $\kappa ' = 8\pi G/c^4$ ,$E_n$
is the stationary state energy in the proper gravitational field
with  the  constant $K$, $\Lambda =\kappa \mu $ , $r_n$ is the
value of the  co-ordinate  $r$,
satisfying  the equilibrium $n$ state in the proper  gravitational
field, $\bar \kappa =\kappa _0\kappa $, $\kappa _0$   is  the
dimensionality constant, $S_{\alpha }=\bar \Psi \gamma _{\alpha }
\gamma _5\Psi $, $\nabla \mu $  is  the  spinor-
coupling covariant derivative independent of twisting, $E'_n$   is  the
energy  state of the particle having a mass $m_n$ (either  free  of
field or being in the external field) and described by the  wave
function $\Psi '$, in the proper gravitational filed with the constant
$G$. The rest notations are generally known in the Gravity Theory.

Equations (\ref{1}) through (\ref{5}) describe the equilibrium states  of  a
particle  (stationary states) in its proper gravitational  field
and determine the localisation region of the field characterised
by  constant  $K$  that  satisfies the  equilibrium  state.  These
stationary states are the sources of the field with the constant
$G$,  and  the condition (\ref{3}) provides matching the solutions  with
the  constants  $K$  and $G$. The proposed model is compatible  with
Quantum  Mechanics principles, and gravitational field with  the
constants $K$ and $\Lambda $ at a certain, quite definite distance specified
by  the  equilibrium state transforms to the  filed  having  the
constant $G$ and satisfying, in the weak field limit, the  Poisson
equation.

A set of equations (\ref{1}) through (\ref{5}), first of all, is of interest
for  the  problem  of stationary states, i.e.,  the  problem  of
energy   spectrum   calculations  for   elementary   source   in
gravitational  field.  Here it seems to  be  reasonable  to  use
analogy with electrodynamics, in particular, with the problem of
electron stationary states in the Coulomb field. Transition from
the   Schr\"{o}dinger  equation  to  the  Klein-Gordon  relativistic
equations  allows  to take into account fine  structure  of  the
electron   energy  spectrum  in  the  Coulomb   field,   whereas
transition  to  the Dirac equation allows to take  into  account
relativistic  fine  structure and  the  energy  level  splitting
associated with spin-orbital interaction. Using this analogy and
appearance  of the equation (\ref{1}), one can conclude that  solution
of  this  equation  without the term $\bar \kappa \bar \Psi 
\gamma _{\mu }\gamma _5 \Psi \gamma _5$ results  in  the  spectrum
similar\footnote{in terms of relativism and removal of degeneracy
by general quantum number} to that of fine structure. As for the
term $\bar \kappa \bar \Psi \gamma _{\mu }\gamma _5
\Psi \gamma _5$, as it  was
already marked in Ref. 1, its contribution is similar to that of
the  term $\bar \Psi \sigma ^{\mu \nu  }\Psi F_{\mu \nu }$  in 
the  Pauli  equation. The  latter  implies  that
solution of the problem of stationary states with twisting taken
into  account  will give total energy-state spectrum  with  both
relativistic fine structure and energy state splitting caused by
spin-twist interaction taken into account. This fact,  being  in
complete  correspondence with requirements of  Gauge  Theory  of
Gravity,  forces us to believe that the above-stated assumptions
on  properties  of  gravitational field in  the  quantum  region
refer,  in  general,  rightly to the  gravitational  field  with
twists.
Due to complexity of solving this problem, we have use a simpler
approximation,   namely:   energy   spectrum   calculation    in
relativistic fine-structure approximation. In this approximation
the  problem of the elementary source stationary states  in  the
proper  gravitational field is reduced to solving the  following
equations:

\begin{eqnarray}
f''+\left(\frac{\nu '-\lambda '}{2}+\frac{2}{r}\right)f'+e^{\lambda}
\left(K_n^2e^{-\nu }-K_0^2-\frac{l(l+1)}{r^2}\right)f=0
\label{6}
\end{eqnarray}

\begin{eqnarray}
-e^{-\lambda }\left(\frac{1}{r^2}-\frac{\lambda '}{r}\right)+\frac{1}{r^2}+\Lambda =
\beta (2l+1)\left\{ f^2 \left[e^{-\lambda }K_n^2+K_0^2+\frac{l(l+1)}{r^2}\right]+
f'^2 e^{-\lambda}\right\}
\label{7}
\end{eqnarray}

\begin{eqnarray}
-e^{-\lambda }\left(\frac{1}{r^2}+\frac{\nu '}{r}\right)+\frac{1}{r^2}+\Lambda =
\beta (2l+1)\left\{ f^2 \left[K_0^2-K_n^2e^{-\nu }+\frac{l(l+1)}{r^2}\right]-
e^{\lambda }f'^2\right\}
\label{8}
\end{eqnarray}

\begin{eqnarray}
\left\{-\frac{1}{2}\left(\nu ''+\nu '^2 \right)-(\nu '+\lambda ')
\left(\frac{\nu '}{4}+\frac{1}{r}\right)+\frac{1}{r^2}(1+e^{\lambda })\right\}_{r=r_n}=0
\label{9}
\end{eqnarray}

\begin{eqnarray}
f(0)=const\ll \infty 
\label{10}
\end{eqnarray}

\begin{eqnarray}
f(r_n)=0
\label{11}
\end{eqnarray}

\begin{eqnarray}
\lambda (0)=\nu (0)=0
\label{12}
\end{eqnarray}

\begin{eqnarray}
\int\limits_0^{r_n}f^2r^2dr=1
\label{13}
\end{eqnarray}
                                                            
Equations (\ref{6})--(\ref{8}) follow from the equations (\ref{14})–-(\ref{15})

\begin{eqnarray}
\left\{-g^{\mu \nu }\frac{\partial }{\partial x_{\mu }}
\frac{\partial }{\partial x_{\nu }}+g^{\mu \nu }\Gamma _{\mu \nu }^{\alpha }
\frac{\partial }{\partial x_{\alpha }}-K_0^2 \right\}\Psi =0
\label{14}
\end{eqnarray}

\begin{eqnarray}
R_{\mu \nu }-\frac{1}{2}g_{\mu \nu }R=-\kappa \left( T_{\mu \nu }-
\mu g_{\mu \nu }\right),
\label{15}
\end{eqnarray}
                                                    
after  substitution of $\Psi $  in the form: $\Psi =f_{El}(r)Y_{lm}(\vartheta ,\varphi )
exp\left(\frac{-iEt}{\hbar}\right)$  and specific  computations
in   the   central-symmetry  field  metric  with  the   interval
determined by the expression \cite{4}

\begin{eqnarray}
dS^2=c^2e^{\nu }dt^2-r^2\left(d\vartheta^2+sin^2\vartheta d\varphi ^2 \right)-
e^{\lambda }dr^2
\label{16}
\end{eqnarray}

he  following  notations are used above: $f_{El}$  is  the  radial  wave
function that describes the states with a definite energy $E$  and
the  orbital  moment $l$ (bellow the indices $l$ are omitted),
$Y_{lm}(\vartheta ,\varphi )$  are spherical functions; $K_n=E_n/\hbar
c$, $K_0=cm_0/\hbar$ , $\beta =(\kappa /4\pi )(\hbar /m_0)$.
The  condition (\ref{9}) determines $r_n$; whereas equations (\ref{10}) through
(\ref{12}) are the boundary conditions and the normalisation condition
for the function $f$ respectively. The general form of the equation
(\ref{9})  is  as  follows: $R(K,r_n)=R(G,r_n)$. In neglect of the proper  gravitational
field with constant $K$, we can re-write this condition as $R(K,r_n)=0$, going
to the equality (\ref{9}).

R.h.s  of  Eqs.  (\ref{7})--(\ref{8}) are calculated on  a  base  of  general
expression for the energy-momentum tensor of the complex  scalar
field:

\begin{eqnarray}
T_{\mu \nu }=\Psi _{,\mu }^+\Psi _{,\nu }+\Psi _{,\nu }^+\Psi _{,\mu }-
\left(\Psi _{,\mu }^+\Psi ^{,\mu }-K_0^2\Psi ^+\Psi \right)
\label{17}
\end{eqnarray}

The  appropriate components $T_{\mu \nu }$ are obtained by summation over the
index  $m$  with  application of the characteristic identities  of
spherical  functions \cite{5} on substituting $\Psi$ =$f(r)Y_{lm}(\vartheta ,\varphi )$
$exp\left(\frac{-iEt}{\hbar}\right)$  to Eq. (\ref{17}).  Even  in
the  simplest approximation the problem of the elementary source
stationary  states  in  the  proper  gravitational  field  is  a
complicated mathematical problem. It is getting simpler  is  one
restricts himself by estimation of the energy spectrum.  Eq. (\ref{6})
can be reduce to the equations \cite{6}:

\begin{eqnarray}
f'=fP(r)+Q(r)z \qquad z'=fF(r)+S(r)z
\label{18}
\end{eqnarray}

This  transition  implies specific choice of $P$,  $Q$,  $F$,  $S$  with
satisfaction of the conditions:
\begin{eqnarray}
P+S+Q'/Q+g=0  \qquad FQ+P'+P^2+Pg+h=0
\label{19}
\end{eqnarray}

where  $g$ and $h$ correspond to Eq. (\ref{6}) written in the form:
$f''+gf'+hf=0$ .  The
conditions  (\ref{19}) are satisfied, in particular, by $P$, $Q$, $F$, $S$
written as follows:
\begin{eqnarray}
Q=1, \qquad P=S=-g/2, \qquad F=\frac{1}{2}g'+ \frac{1}{4}g^2-h
\label{20}
\end{eqnarray}

Solutions of the set  (\ref{18}) are the functions \cite{6}:
\begin{eqnarray}
f=C\rho (r)sin\vartheta (r) \qquad z=C\rho (r)cos\vartheta (r)
\label{21}
\end{eqnarray}

where  $C$  is  an  arbitrary constant, $\vartheta (r)$  is the  solution  of  the
equation:
\begin{eqnarray}
\vartheta '=Qcos^2\vartheta +(P-S)sin\vartheta cos\vartheta -Fsin^2\vartheta ,
\label{22}
\end{eqnarray}

and $\rho (r)$ is determined by the formula
\begin{eqnarray}
\rho (r)=exp\int\limits_0^r\left[Psin^2\vartheta +(Q+F)sin\vartheta cos\vartheta +
Scos^2\vartheta  \right]dr.
\label{23}
\end{eqnarray}

In  this  case, the form of solution presentation in  polar  co-
ordinates allows to determine zeros of the functions $f(r)$  at $r=r_n$,  with
correspondent  values  of $\vartheta =n\pi $  ($n$ is an integer).  As  one  of  the
simplest approximations for $\nu $, $\lambda $ , let's choose the dependence:
\begin{eqnarray}
e^{\nu }=e^{-\lambda }=1- \frac{\tilde{r_n}}{r+C_1}+ \Lambda (r-C_2)^2+C_3r
\label{24}
\end{eqnarray}

where $\tilde{r_n}=\frac{2Km_n}{c^2}=\frac{2KE_n}{c^4}=
\left(\frac{2K\hbar}{c^3}\right)K_n$, $C_1=\frac{\tilde{r_n}}
{\Lambda r_n^2}$ , $C_2=r_n$ , $C_3=\frac{\tilde{r_n}}{r_n(r_n+C_1)}$

Earlier  the  estimate for $K$ was adopted  as $K\approx 1.7\times 10^{29}$
$ Nm^2kg^{-2}$.  If  one
assumes that the observed value of the electron rest mass $m_1$  is
its   mass  in  the  ground  stationary  state  in  the   proper
gravitational  field,  then $m_0=4m_1/3$. From dimensionality  reasoning  it
follows that the coupling energy is determined by the expression
$\left(\sqrt{K}m_0 \right)^2/r_1$=$0.17\times 10^6\times 1.6\times 10^{-19}$J,
 where $r_1$ is the electron classic radius. Then we obtain  the
following  estimate: $K\approx 5.1\times 10^{31} Nm^2kg^{-2}$, which 
 is  used  later  as  the
initial  one. Discrepancies in the estimates for $K$, obtained  by
various  ways,  are  quite  admissible,  still,  being  not   of
catastrophic  character. From the fact  that $\mu $  is  the  electron
energy  density  it  follows: $\mu $= $1.1\times 10^{30}J/m^3$, 
$\Lambda =\kappa \mu =4.4\times 10^{29}$ $ m^{-2}$.  From  Eq.  (\ref{22})  it
follows\footnote{with the equation for $f(r)$ taken into account}:
\begin{eqnarray}
2\vartheta '=(1-\bar F)+(1+\bar F)cos2\vartheta \approx (1-\bar F)
\label{25}
\end{eqnarray}

where $\bar F=\frac{1}{2}\bar g'+ \frac{1}{4}\bar g^2-\bar h $,
 $\bar g=r_n\left(\frac{2}{r}+\frac{(\nu '-\lambda ')}{2}\right) $ ,
 $\bar h=r_n^2e^{\lambda}\left(K_n^2e^{-\nu }-K_0^2-\frac{l(l+1)}{r^2}\right)$.

Integration  of Eq. (\ref{25}) and substitution of $\vartheta =\pi n$, $r=r_n$ 
results  in  the dependence of $K_n$ on $r_n$
\begin{eqnarray}
-2\pi n=-\frac{7}{4}-\frac{r_nK_n^2}{\Lambda ^2}\sum \limits_{i=1}^3
\left\{ A_i\left[\frac{(r_n+\alpha _i)^2}{2}-2\alpha _i(r_n+\alpha _i)+
\frac{\alpha _i^3}{(r_n+\alpha _i)}+2C_1(r_n+\alpha _i)+ \right. \right.\nonumber\\
\left. \left. +2C_1\frac{\alpha _i^2}{r_n+\alpha _i}+\frac{C_2^2\alpha _i}{r_n+\alpha _i}\right]+
B_i\left[(r_n+\alpha _i)+\alpha _i^2\frac{1}{r_n+\alpha _i}+\frac{2C_1\alpha _i}
{r_n+\alpha _i}-\frac{C _2^2}{r_n+\alpha _i}\right]\right\}+ \nonumber\\
+\frac{K_0^2r_n}{\Lambda ^2}\sum \limits_{i=1}^3 A'_i(r_n+\alpha _i)+
\frac{r_nl(l+1)}{\Lambda }\left[d_1r_n-\frac{C_1d_2}{r_n}+\sum \limits_{i=1}^3
\alpha _i (r_n+\alpha _i)\right]- \nonumber\\
-\frac{K_n^2r_n}{\Lambda ^2}\left\{\sum \limits_{i=1}^3
\left[2\alpha _i^2A_i-2\alpha _iB_i-4C_1A_i\alpha _i+2C_1B_i+C_2^2A_i+
\frac{K_0^2\Lambda A'_i}{K_n^2}(\alpha _i-C_1)- \right. \right. \nonumber\\
-r_n^2\Lambda l(l+1)\alpha _i
\left. \left. (C_1-\alpha _i)\right]ln(r_n+\alpha _i)-r_n\Lambda ^{-1}l(l+1)(d_2+C_1d_1)lnr_n\right\}
\label{26}
\end{eqnarray}

The  coefficients  entering Eq. (\ref{26}) are the factors  at  simple
fractions  in  the  polynomial  expansion  needed  for  equation
integration, and $\alpha _i\sim K_n$, $d_2\sim A_i\sim r_n^{-5}$, $B_i\sim r_n^{-4}$, 
$A'_i\sim r_n^{-2}$, $\alpha _i\sim r_n^{-4}$, $d_1=r_n^{-4}$. There exists the condition
(\ref{9})  (or the equivalent condition $exp \; \nu (K,r_n)=1$, used for this approximation) in order
to  eliminate $r_n$  from Eq. (\ref{26}). However, direct application  of
this   condition  will  make  the  expression  (\ref{26})  still  more
complicated.  And one can readily notice that $r_n\sim 10^{-3}r_{nc}$, where $r_{nc}$
is  the Compton  wavelength of a particle of the mass $m_n$, and, hence, 
$r_n\sim 10^{-3}K_n^{-1}$.
The  dependence (\ref{26}) itself is rather approximate; nevertheless,
its  availability,  in  spite  of  the  approximation  accuracy,
implies  existence  of the energy spectrum,  being  sequence  of
particle  self-interaction with its proper  gravitational  field
within  the  range $r\leq r_n$, where mutual compensating  actions  of  the
particle  and  the  field  take  place.  With  $l=0$ the  approximate
solution  (\ref{26}), with the relation between  $r_n$ and $K_n$ taken  into
account, has a form:
\begin{eqnarray}
E_n=E_0\left(1+\alpha e^{-\beta n}\right)^{-1},
\label{27}
\end{eqnarray}

where $\alpha =1.65;\; \beta =1.60$

The  dependence (\ref{27}) is specified on a base of the fact that the
observed value of the electron mass in rest is the value of  its
mass in the grounds stationary state in the proper gravitational
field, and $r_1=2.82\times 10^{-15} m$, $K_1=0.41\times 10^{12} m^{-1}$
result in the accurate zero of the function,
by definition of the numerical values for $K$ and $\Lambda $.

Thus,  the given numerical estimates for the electron show  that
within  the range of its localisation, with $K\sim 10^{31}N m^2 kg^{-2}$
  and $ \Lambda \sim 10^{29}m^{-2}$,
the  spectrum  of stationary states in the proper  gravitational
field exists. .The numerical value of $K$ is, certainly, universal
for  any elementary source, whereas the value for $\Lambda $ is determined
by the elementary source mass in rest. The distance at which the
gravitational  field with the constant $K$ is  localised  is  less
than  the Compton wavelength, comprising for electron the  value
of  an  order  of its classical radius. At the distances  larger
than  this one, the gravitational field is characterised by  the
constant $G$, i.e., correct transition to Classical GTR holds.

From  Eq.  (\ref{27}), roughly, the numerical values of the stationary
state energy follow; $E_1=0.511\; MeV$, $E_2=0.638\; MeV$, $\ldots$
$E_{\infty }=0.681\; MeV$. The quantum transitions
over  stationary states, allowed by selection rule, must  result
in  the gravitational emission characterised by constant $K$.  The
natural  widths  of  transition energies in this  spectrum  will
comprise  from  $ 10^{-9}$ to $10^{-7} eV$. A small value  of  the  energy
level  width,  compared to the electron energy  spread  in  real
conditions,  explains  why  the gravitational  emission  effects
aren't  observed  as  by-passers,  e.g.,  in  the  processes  of
electron beam bremsstrahlung. If one manages, anyway, satisfying
the  conditions for excitation of gravitational emission,   then
availability   of   large  constant  of  gravity   must   effect
considerably  on  a  state  of the emitting  system,  i.e.,  the
observed  effects of gravitational field may turn out  far  from
the traditionally assumed ones.

\section{Energy of Gravitational Field As Hidden Energy of
Universe}

As  is  known \cite{7}, in terms of the Robertson-Worker  metric  the
fundamental  equations  of Dynamical Cosmology  are  written  as
follows\footnote{in this system of measurement $c$=1}:
\begin{eqnarray}
3\ddot{a}=-4\pi G(\rho +3p)a
\label{28}
\end{eqnarray}

\begin{eqnarray}
a\ddot{a}+2\dot{a}^2+2k=4\pi G(\rho -p)a^2
\label{29}
\end{eqnarray}

\begin{eqnarray}
\dot{p}a^3=\frac{d}{dt}a^3(p+\rho )
\label{30}
\end{eqnarray}

\begin{eqnarray}
p=p(\rho )
\label{31}
\end{eqnarray}

Equation (\ref{30}) in Cosmology is treated as the energy conservation
law;  however,  this  assertion is to be refined,  because  this
equation  is  sequence of the Bianki identities $\left(R^{\mu \nu }-
\frac{1}{2}g_{\mu \nu }R \right)_{;\nu }\equiv T_{;\nu }^{\mu \nu }=0$ .
 As  for  the equations  $T_{;\nu }^{\mu \nu }=0$,  it's well known that
 they aren't the equations  of
energy-momentum conservation. In accordance with  stated  above,
the particle stationary fields in the proper gravitational field
with  the  constant  $K$  serve  a  role  of  the  source  of  the
gravitational  field of an isolated particle  with  the  gravity
factor $G$. The gravitational field energy of a particle with  the
constant  $G$  and  the mass $m$ equals the difference  between  the
energy  of a particle having the mass in rest $m_0$ (being a source
of  the  field  with the constant $K$) and energy  of  a  particle
having the mass in rest $m$. Then, evidently, by definition of the
gravitational field energy-momentum tensor $t^{\mu \nu }$, it follows that the
law  of  energy-momentum conservation for a particle having  the
mass  $m$  in the gravitational field with the constant $G$ has  the
form: $(T^{\mu \nu }(m)+t^{\mu \nu })_{,\nu }$. In terms of the
Robertson-Worker metric $\Gamma _{ii}^{\mu }=0$, and, hence, the
appropriate  identity  by  Bianki  is  the  energy  conservation
equation,  provided  in  the energy-momentum  tensor  for  ideal
liquid $p$  is  replaced by $p_{\epsilon }=p+\rho _{g}$, where
$\rho _{g}$ is the energy density  of  the
gravitational  field.  And  it is the  focus  of  refinement  of
Dynamical  Cosmology  equations (\ref{28}) through  (\ref{31}),  being  very
significant because it leads to the relation $\rho _0/\rho _c=2/(1+3\alpha _g)$,
where $\rho _0$  is  the substance  energy  density, $\rho _c$ is 
the  critical  value  of  the substance density, $\alpha_g=\rho _g/\rho$,
and the retarding parameter $q_0$ is set to be
equal  to  unity. The numerical value $\alpha _g$, e.g., for  an  electron,
equals 1/3, whereas for nucleons, being in overwhelming majority
in  the  Universe, with its rather complicated  structure  taken
onto  account, $\alpha _g$ can have the meaning quite close  to  unity  and
even a little bit higher. On the other hand, the numerical value $\rho _0/\rho _c$
In Cosmology is estimated as being within the range from $0.03$ to
$0.06$,  and  higher  \cite{8}.  In the course  of  refinement  of  the
observed  meaning  for $\rho _0$ and calculated $\alpha _g$ one  should,  seemingly,
expect  that  the  values  of  the  observed  ratio $\rho _0/\rho _c$  and   that
calculated according to the formula $2/(1+3\alpha _g)$ come closer.
Thus,  in  terms of quanta, gravitational interaction  leads  to
existence of considerable fraction of energy in the form of that
of  the  gravitational field, i.e., to presence of  hidden  mass
which  should  be taken into consideration in the  equations  of
Dynamical Cosmology.

\section{Compression of High-Temperature Plasma System by Emitted
Gravitational Field}

In  the dense high-temperature plasma the gravitational emission
can  occur as a result of electron bremsstrahlung at the  nuclei
of  ion  components, i.e., the gravitational  emission  in  this
plasma,   as   well  as  the  electromagnetic   one,   are   the
bremsstrahlung  emission.  Numerical values  of  the  transition
energies  for  stationary states of an electron  in  the  proper
gravitational  field are, approximately, within the  range  from
$127$  to  $171 keV$, with the transition energy width value  of  an
order  of $10^{-7} eV$, and that's why the gravitational emission  in
plasma  may  occur  only  if  its density  and  temperature  are
sufficiently high. The recoil energy, as one can readily verify,
comprises tens of electron-volts, i.e., resonance absorption  of
gravitons  on  nuclei  is absent. As the Compton  scattering  of
gravitons  on nuclei is insignificant, then it is sufficient  to
consider  the Compton graviton scattering from electrons,  being
crucial for plasma state in the emitted gravitational field.
The  frequency of elastic electron-ion collisions in  plasma  is
determined by the well-known expression \cite{9}
\begin{eqnarray}
\nu _{ii}=\frac{1}{6\pi \epsilon _0^2\sqrt{2\pi m}}\frac{e^2e_i^2n_i}
{(kT_e)^{3/2}}L_e
\label{32}
\end{eqnarray}

where $\epsilon _0$, $T_e$, $k$, $n_i$, $L_e$ are, respectively, the electric constant,
the   electron  temperature,  the  Boltsman  constant,  the  ion
component concentration and the Coulomb logarithm.

In  the  simplest  approximation, in view of estimation  of  the
value  of  the electrostatic (intra-plasma) field in plasma  one
can use the expression \cite{10}:
\begin{eqnarray}
E=\frac{e_i}{4\pi \epsilon _0(4\pi n/3)^{-2/3}}
\label{33}
\end{eqnarray}

where $n$ is the total concentration of plasma.
For  small  values  of the internal plasma field  intensity  the
bremsstrahlung   gravitational  emission   (as   well   as   the
electromagnetic one) is determined by the appropriate energy  of
thermal random motion of electrons. As for large values  of  the
internal plasma field intensity (being considered below), it  is
the   value   of  field  intensity  crucial  in  excitation   of
gravitational  emission.  In fact,  the  average  value  of  the
kinetic  energy  of an electron subject to acceleration  in  the
internal  plasma  field, in accordance with  Eqs.  (\ref{32})  through
(\ref{33}), is determined by the expression:
\begin{eqnarray}
\frac{M}{2}\left[\frac{ee_i(4\pi n)^{2/3}(T_ek)^{2/3}6\pi 
\epsilon _0^2\sqrt{2\pi m}}{4\pi \epsilon _0\cdot 2M\cdot e^2e_i^2n_iL_e}
\right]^2=E_K
\label{34}
\end{eqnarray}

where $M$ is the electron relativistic mass.
Significant  increase in the energy density of stationary  sates
of  an  electron  in  its proper gravitational  field  initiates
beginning  with $\sim 167 keV$, i.e., the condition for excitation  of
plasma  electron gravitational emission under valuable  internal
plasma field intensity will have the form:
\begin{eqnarray}
E_K\geq 1.67\times 1.6\times 10^{-14}J.
\label{35}
\end{eqnarray}

Thus, if the plasma parameters reach the values that satisfy the
condition  (\ref{35}), then, beginning with this moment  of  time,  in
plasma  considerable number of elementary acts of  gravitational
emission take place, although the average value of the energy of
thermal   random  motion  of  electrons,  in  neglect   of   its
acceleration by internal plasma field, is lower, at least, by an
order   of   magnitude.  The  process  of   the   bremsstrahlung
gravitational   emission,   as  well   as   the   bremsstrahlung
electromagnetic emission, is accompanied by electron scattering.
Hence,  the  cross-section  of the bremsstrahlung  gravitational
emission  is  represented  as  the product  of  probability  for
graviton  emission (as the first-order process) and the electron
elastic scattering cross-section. This fact serves as the ground
for   application   of  the  expression  (\ref{32})  when   performing
approximate   estimation  of  the  excitation   conditions   for
generation of gravitational emission of plasma electrons subject
to acceleration in the internal plasma field.
Expressions    for    the   cross-section   of    bremsstrahlung
electromagnetic emission, in general case, are very complicated.
However, one can use the Born approximation, as the Born  cross-
section form is rather simple. And although the energy range  of
interest  is  far beyond the Born approximation, its application
will  provide  us useful qualitative information concerning  the
electromagnetic emission intensity. The bremsstrahlung  emission
cross-section in this approximation is as follows \cite{11}:
\begin{eqnarray}
\sigma _e=\frac{8}{3}\frac{r_0^2z^2}{137}\frac{mc^2}{E_0}
ln\frac{(\sqrt{E_0}-\sqrt{E_0-\varepsilon })^2}{\varepsilon }
\label{36}
\end{eqnarray}

where  $z$  is  the index of an ion component, $E_0$ is the  electron
primary energy, $\varepsilon $  is the energy of an emitted photon, $r_0$ is  the
electron classic radius.
The energy emitted by the unit of plasma volume per the unit  of
time  within  the  frequency range   $d\varepsilon $  is  determined  by  the
expression:
\begin{eqnarray}
dQ_e=\sigma _en_en_i\sqrt{\frac{2E_0}{m}f_{E_0}}
\label{37}
\end{eqnarray}

where $f_{E_0}$  is the function of electron distribution over the values
of $E_0$.

In   Ref.  \cite{12}  the  expression  (\ref{37})  is  integrated  for  the
Maxwellian distribution for values of $E_0$ and $\varepsilon $, and, as a result,
the following formula for $Q_e$ is obtained:
\begin{eqnarray}
Q_e=\frac{32}{3}\frac{z^2r_0^2}{137}mc^2n_en_i\sqrt{\frac{2kT_e}{\pi m}}
\label{38}
\end{eqnarray}

where $T_e$ temperature of the electron component.

For  approximate assessment of the bremsstrahlung  gravitational
emission  intensity,  one  can  use   the  expression  (\ref{38}),  on
replacing  $r_0$  by  $r_g$,  calculated  by  the  formula $r_g=2Km/c^2$,  being
correspondent to a replacement of the electric charge $e$  by  the
gravitational  charge $m\sqrt{K}$.  Then the bremsstrahlung  gravitational
emission   intensity  of  the  plasma  electron   component   is
determined by the expression:
\begin{eqnarray}
Q_g=\beta \cdot Q_e
\label{39}
\end{eqnarray}

where $\beta =0.16$ for the obtained numerical value of $K$.

For  sake  of simplicity, let the region occupied by  plasma  be
spherically symmetric one, having the radius $r_0$ and  the  radius
of gravitational emission region $r_{0g}$ determined by the condition
(\ref{35}).   After  excitation  of  gravitational  emission,  it   is
intensified with the growth of the internal plasma field at  the
expense  of  the  increase in the number of emission  elementary
acts  and, probably, in the size of the emission region as well.
Transition  from  intensification  to  generation  occurs,   the
increase in emission intensity is higher than its loss, and this
is the case only if emission is locked up  within plasma.

When  creating  high-temperature  plasma  states  at  laboratory
conditions, they use, first of all, light gases, because maximum
temperatures are accessible for them with minimum energy pumping-
in.  In  particular,  overwhelming majority  of  experiments  on
investigation  of  small-duration $z$-pinch-type pulse  discharges
were performed  with application of deuterium \cite{13}. Increases in
the  plasma  temperature  and density in  pulse  discharges  are
related  to  plasma  compression  by  the  magnetic  field.  The
electron plasma frequency of compressed plasma is determined  by
the known expression $\omega _{L_e}=\sqrt{e^2n_e/4\pi \epsilon _0m}$.
Evidently, for the emission region having
the  radius $r_{0g}$  the  condition $r_{0g}<r_0$  is satisfied.
Let's  adopt  that $\omega _{0g}$
denotes  the  frequency  of the emitted gravitational  field  at 
$r=r_{0g}$ (i.e.,  along the boundary of the emission region).  Then,  with
analogy between the impacts of electromagnetic emission and  the
gravitational  one on plasma electrons taken into consideration,
the condition for confined emission in plasma can be written  as
follows:
\begin{eqnarray}
\omega _g(r_0)=\sqrt{e^2n_e(r_0)/4\pi \epsilon _0m}
\label{40}
\end{eqnarray}

The  value of $\omega _g(r_0)$ is determined by graviton Compton scattering from
plasma  electrons at the distanced within the range from $r_{0g}$  to $r_0$ ,
i.e., is found as a result of solution of the equations
\begin{eqnarray}
d\omega _g/\omega _g=g_{rr}^{-\frac{1}{2}}\sigma _gn_e(r)dr
\label{41}
\end{eqnarray}

where $\sigma _g$  is  the cross-section of the graviton Compton scattering
from  plasma electrons, and the initial condition for $\omega _g$  has  the
form: $\omega _g(r_{0g})=\omega _{0g}$.

Satisfaction  of the condition (\ref{40}) means presence  of  positive
feedback in the system, i.e., this condition leads to generation
of gravitational emission  in plasma subject to compression.
It  follows from simple qualitative grounds based on analysis of
the function (\ref{34}) that the conditions for gravitational emission
excitation  in  plasma are accessible to higher extent  for  the
plasma composed of, at least, two components: $\alpha _1z_1+\alpha _2z_2$,
where $\alpha _1$ and $\alpha _2$  are
the  weight fractions of the light ion component (hydrogen)  and
the  heavy one (carbon, oxygen, nitrogen). This follows from the
fact that the internal plasma field intensity grows sharply when
the  condition $n_e > n_i$ is met; i.e. application of  multi-charge
ion   component  is  required.  The  condition  (\ref{34})  determined
specific  size  of  the region where gravitational  emission  of
plasma  occurs as a result of plasma compression for a specified
plasma  composition.  Compared to the case  of  purely  hydrogen
plasma, larger energy pumping-in to the two-component plasma  is
required in order to keep unchanged the values of $n_e$, $n_i$, $T_e$
in the  peripheral  region as the value of $\alpha _2$ is getting  higher;  it
means  that upper restriction to $\alpha _2$  exists. On the other hand,  it
follows  from Eqs. (\ref{40}) and (\ref{41}) that the less is $\alpha _2$,
the  greater
is  the distance at which gravitational emission suppression  in
plasma  takes  place. Hence, the value of $\alpha _2$ is to  be  such  that
emission suppression takes place for $r>r_{0g}$ and, at the same time, for
$r\leq r_0$,  leading to occurrence of restriction on $\alpha _2$ from below. The role
of  a heavy component is not limited to producing conditions for
excitation   of  gravitational  emission  but  is  crucial   for
producing  conditions for confined emission  in  plasma,  as  it
follows  from (\ref{40}) - (\ref{41}). In fact, the value of $\sigma _g$ is significantly
greater  than  the  similar value of the  cross-section  of  the
Compton  scattering of emission on ions. This implies  that  the
efficacy  of  confinement of emission in plasma as well  as  the
conditions  for  its  excitation  are  determined  also  by  the
condition $n_e > n_i$, i.e. it can take place in plasma with  multy-
charge ions only.

When  comparing, on a base of Eq. (\ref{39}), the gas kinetic pressure
in  plasma for the plasma parameters that satisfy the excitation
conditions  for gravitational emission and the pressure  of  the
emitted  gravitational field, one can be convinced that  in  the
time  interval  of  an  order of $t_g=10^{-6} s$ after emission  commencement
these pressures are of the same order. Hence, approximately,  in
$t_g=10^{-6} s$  after  commencement of gravitational emission  in  compressed
plasma  the conditions needed for its confinement by the emitted
gravitational  field  (i.e., the states  of  plasma  hydrostatic
equilibrium in the emitted gravitational field) are achieved  if
the   latter  is  confined  in  plasma.  In  addition   to   the
gravitational   emission,  the  bremsstrahlung   electromagnetic
emission occurs over the entire region of compressed plasma;  so
only  a  part  of  the  emission  is  confined  in  the  region,
practically,  common  to gravitational emission.  The  remaining
part of the emission transforms to plasma thermal loss, and  the
emission  spectrum  corresponds to the  thermal  energy  of  the
electron random motion. The confined part of the electromagnetic
bremsstrahlung emission returns its energy to plasma as a result
of  collisions  and  it  almost  doesn't  take  part  in  plasma
compression  ,  in  contrast to the emitted gravitational  field
having  the pressure gradient that coincides precisely with  the
gas   kinetic  pressure  gradient.  Thus,  with  the   satisfied
conditions   for   excitation  of  gravitational   emission   in
compressed  two-or-more-component  plasma,  intensification  and
generation of the emission occur. It should be stressed that   a
heavier  component  is needed for confinement  of  gravitational
emission.

The  presented  analysis  allows to draw  some  preliminary  and
rather impressive conclusions as follows:

 1.   The fundamental property of the emitted gravitational field
   consists in the fact that it compresses the emitting system as
   the emission intensifies.

 2.   A system in which the gravitational emission is excited and
   intensified begins to operate as the quantum generator with the
   operation output consisting in achievement of the hydrostatic
   equilibrium states in the emitted gravitational field, rather
   than the emission release out of the system.

As  is mentioned above, the well-known way for production of the
dense  high-temperature states in plasma is its  compression  by
magnetic field, being especially effective if so-called  "plasma
focus"-type installations (PF) are applied. With application  of
the PF installation, the quite high energy capacity but not very
stable  states of plasma are produced, provided a single gas  is
used  as  the operational substance.. Addition of a heavier  gas
(e.g.,  xenon)  to  light operational gaseous agent  results  in
occurrence  of  the compression regime, in which plasma  at  the
final  stage  serves as a source of the intense  X-ray  emission
\cite{14}.  A  well-known  technique of plasma  generation  for  high
atomic  number  elements is application of  vacuum  diodes  with
initiated  break-down  in  the  inter-electrode  gap.   In   the
discharges  like  this local high-temperature plasma  formations
(LHPF)  are observed. Its nature can't be explained by pinch  in
magnetic  lines of force. The break-down distinguishing  feature
(similar  to the case of "PF" at the X-ray emission  regime)  is
presence  of  multi-charge ions, i.e., excess  of  the  electron
component  concentration compared to the ion one. In  accordance
with  stated  above,  occurrence of LHPF  can  be  explained  by
compression  of  the  breakdown local  regions  exerted  by  the
emitted gravitational field, because, owing to presence of multi-
charge  ions,  the condition for reinforcement of  gravitational
emission is satisfied.
Let's now go back to the pulse strong-current discharges in  the
PF-type installations. The PF regime (with xenon admixture) with
the  X-ray  emission  occurring  at  the  final  stage  can   be
considered  as the intermediate stage between the  dense  plasma
unstable focus regime and the stage at which reaching the  state
of  plasma  hydrostatic equilibrium in the emitted gravitational
field is possible. The fact that the phenomenon like this hasn't
been  observed  experimentally is associated  with  imperfection
(non-optimum procedures) of the experiments concerning both  the
numerical  ratio  of  light-to-heavy components  and  the  heavy
component  index. Hence, in view of experimental record  of  the
fact that the state of plasma hydrostatic equilibrium is reached
in the emitted gravitational field in the pulse strong-current $z$-
pinch-type  discharges,  special  experiments  with  binary  gas
mixtures like \emph{hydrogen + carbon/oxygen/nitrogen} are needed.  And
the optimum condition related to the plasma equilibrium state in
the  emitted  gravitational  field is,  evidently,  the  minimum
recorded integral intensity of the above-thermal portion of  the
electromagnetic emission under the growth of the energy  pumped-
up to discharge.

As  for  the  binary  mixtures,  application  of  a  composition
containing $80\%$ of hydrogen and $20\%$ of carbon isotope 
$\mathbf{{}_{6}C^{12}}$  seems
to  be rather attractive. As is known, the nuclear transmutation
chain involving carbon isotope $\mathbf{{}_{6}C^{12}} $ is called
 carbon cycle.  The
carbon  cycle  results  in a conversion  of  four  protons  into
$\alpha $  particle followed by $26.8 \; MeV$ energy output, i.e. the  carbon
chain  concludes  with  a thermonuclear  fusion  reaction.  This
composition of the initial gas mixture is applicable in terms of
accessibility  of the hydrostatic equilibrium  states.  But  the
same  composition is applicable for the controlled thermonuclear
fusion,  and  this  fact is to be of interest for  experimenting
correspondingly on pulse strong-current discharges.

\section{Gravitational Emission Accompanying $\beta $ -Decay}
The   analysis   performed  above  shows  that   (provided   the
assumptions  on  the  quantum properties  of  the  gravitational
impact  are valid) the gravitational emission can be excited  in
the    dense   high-temperature   plasma;   however,    emission
intensification  results in compression of the emitting  system.
Hence, as the gravitational emission increases, only sequence of
the  gravitation  emission  will be  observed  rather  than  the
emission itself\footnote{Numerous studies devoted to recording the gravitational waves
\cite{15} aren't successful, being based on linear approximation
which can't takre place in the case of large value of the
emotted field gravitational constant.}. This fact doesn't allow to support validity of
the   assumption  stated  above  and,  moreover,  to   determine
numerical   characteristics  of  the  particle  stationary-state
spectrum  in  the  proper gravitational field.  In  terms  of  a
principal  experimental test, electrom is  the  most  applicable
object  having estimation (though rough) of its stationary-state
spectrum  in the proper gravitational field. Also the  processes
exist (such as natural/artificial decay of elementary particles)
that  have  nothing  common to emission growth  and  where  pure
gravitational   emission  can  be  observed.  The   essence   of
observation  for the elementary particle decay process  consists
in  the  point that (similar to the case of chemical  reactions)
the particles produced as a result of decay can be in an excited
state  with respect to the ground stationary state in its proper
gravitational fields. In this respect, the $\beta $ decay processes seem
to  be  rather  attractive, because its  experimental  recording
procedures are quite perfect. As is known, asymmetry of  emitted
electrons,  considered  as caused by parity  nonconservation  in
weak interactions, is typical for $\beta $ decay \cite{16}. Nevertheless,  on
a  base  of  healthy grounds, this fact should  be  regarded  as
anomaly,  because  in  other types of  fundamental  interactions
parity  is conserved. The $\beta $ asymmetry in the angular distribution
of  electrons  is  recorded  in the experiments  with  polarised
$\mathbf{{}_{27}Co^{60}}$ nuclei having the $\beta $ spectrum which is characterised by the
energies of an order of several $MeV$. If in the process of $\beta $ decay
production  of excited electrons takes place, then, in  addition
to the decay scheme:
\begin{eqnarray}
n\rightarrow p+e^-+\tilde{\nu }
\label{42}
\end{eqnarray}

also the following scheme will be realised:
\begin{eqnarray}
n\rightarrow p+(e^*)^-+\tilde{\nu }\rightarrow e^-+\tilde{\gamma }+\tilde{\nu }
\label{43}
\end{eqnarray}

where $\tilde{\gamma }$ is the graviton.

The decay described by Eq. (\ref{43}) is confined by the energy values
of  an  order of $1 MeV$ (in rough approximation), with  the  fact
taken into account that the difference between the lower excited
level of the electron energy (in its proper gravitational field)
and  the ground state  equals $\sim 130\; keV$, and by the character  of
the $\beta $  spectrum  as well. Hence, decay of the $\mathbf{{}_{27}Co^{60}}$
 nuclei  can
occur  with the same probability following both the scheme  (\ref{42})
and  the  scheme (\ref{43}). For light nuclei (e.g.,$\mathbf{{}_{1}H^{3}}$)
the $\beta $  decay
can  be  implemented only by the scheme (\ref{42}). And it is emission
of the graviton by an electron in magnetic field (with potential
electron  energy  level splitting in magnetic field  taken  into
consideration)  can  lead  to the $\beta $  asymmetry  of  the  electron
angular   distribution.  If  it's  not  true,  then  for   light
$\beta $-radioactive nuclei the phenomenon of $\beta $ asymmetry isn't observed.
It  means  that $\beta $ asymmetry of the element angular distributions,
treated  as  parity nonconservation, is the consequence  of  the
electron  gravitational emission, and, hence, the lower boundary
for $\beta $ asymmetry of the $\beta $ decay must exist.

\section{Conclusion}
                                
 1.   The approximation for the Einstein relativistic gravity has
   been  considered  in  which the values of  the  gravitational
   constant  and  the  constant $K$ for the region  of  elementary
   particle  localisation are specified such way that stationary
   states of particles in the proper gravitational field occur, and
   the particle stationary states themselves are the sources of the
   field with the Newtonian gravitational constant $G$.

 2.    Presence of the Universe hidden energy in the form of the
   gravitational fields is the result of existence of the particle
   stationary states in the proper gravitational field, and this
   fact is to be taken into account in the equations of dynamical
   Cosmology, in accordance with significance of the hidden energy
   for Universe life at present.

 3.   Accessibility of the hydrostatic equilibrium states in the
   dense high-temperature plasma in the emitted gravitational field
   is   potential  consequence  of  the  properties  of  quantum
   gravitational interaction. This fact can be examined  at  the
   experiments with pulse strong-current $z$-pinch-type discharges.

 4.    Presence  of the lower coupling energy boundary  for  the $\beta $
   asymmetry of the electron angular distribution in the $\beta $ decay may
   be  considered  as direct confirmation of the fact  that  the
   gravitational emission of electrons exist, in particular, in the $\beta $
   decay, exists.

\title{\textbf{Can a link between asymmetry of the electron angular
distribution in the $\beta $ decay processes and controlled thermonuclear
fusion exist? }}

\maketitle

\begin{abstract}
Comments on a manuscript of the article 'Anomalies In The $\beta $ decay Processes And The Pulse Strong-
Current Discharges As Consequence Of Electron Gravitational Emission'

\end{abstract}

1. The equations of the Relativistic Theory of Gravity worked
out by Einstein from the variation principle are the most general form
of such equations satisfying the general covariance principle.
The requirement of correspondence to the Newtonian Classic Theory of
Gravity has lead Einstein to the General Theory of Relativity.
And that corresponds to the equations of the Relativistic Theory of Gravity
with $\Lambda =0$, while the coupling constant $K$ is equal to the gravitational
constant $G$. Beginning with seventies, it became clear [1] that in quantum
region the numerical value of the constant $G$ isn't compatible with the 
principles of Quantum Mechanics. In a number of papers [1] (also in [2])
it was shown that in quantum region the coupling constant $K$ is more
applicable, while $K\cong 10^{42}G$. So the problem of quantum-level
generalisation of relativity equations was reduced to matching the numerical
values of gravity constants in quantum and classic regions to each other.
The article does not list the various attempts to match the quantum and
classic regions within the frames of the Relativistic Theory of Gravity
(including those applying scalar-tensor theories). These attempts are
discussed in [1] and other more recent publications: 
there are no tangible results.

2. In this work the mentioned problem is being solved by the suggested
theorem-like proposal followed by its rather approximate proof. This proof
can be criticized for its approximate character (an exact proof in 
explicit analytic form is, seemingly, unobtainable due to non-linearity of
equations). However, one cannot reject the suggested theorem because both on
a base of physics and mathematics its result exactly matches the principles
of the Relativistic Theory of Gravitation with the ones of the Quantum 
Mechanics. In effect, the equations of the Relativistic Theory of Gravitation
which incorporate the $\Lambda $-term and coupling constant $K = 10^{42}G$ take place in
quantum region. The limiting transition implied in the suggested proposal
subsequently leads to the equations of the General Theory of Relativity in
classic region. A semiquantum description applied in the work certainly does
not always give the true picture of the quantum world, however such
description does not considerably misrepresent it either. In this very 
case the semiquantum description is used only as an approximation to obtain
estimated numerical values of energy levels of electron in the strong
gravitational field. The manuscript specifically treats such approximations
to be quite rough though giving picture of the subject of study.
The principal points in this case are a possible existence of the energy
levels of electron in the strong gravitational field and primary estimation
of these levels though made by quite rough means.

3. The obtained quantum-level generalization of relativity equations
logically solves the problem of a hidden mass in the Universe, which is
briefly mentioned in the manuscript. Yet the quantity $\alpha _g$ equal to $1/3$ can
be securely used for electrons only. It is known that application of the
dependence $E_0=\frac{e^2}{r_0}=mc^2$  results in the expression for
the momentum $P_i=\frac{4}{3}\frac{E_0}{c^2}\nu _i$  that differs by
the factor of $4/3$ from the correct expression for the momentum of a particle
with mass $m=\frac{E_0}{c^2}$. This very circumstance proves the accuracy of the numerical
value of $\alpha _g = 1/3$ for an electron, because the "extra" part of the energy
is bound (bound part of the energy in the form of the gravitational field
energy). As for other particles, at the current level of calculations one
can speak only of the rough estimation of $\alpha _g$ value, as it is stated in the
manuscript.

4. There is an unexpected conclusion that a system in which the gravitational
emission is excited begins to operate as a quantum generator compressing the
system by emission, rather than releasing emission out of the system. This
means that the gravitational waves must not be observed in nature (except for 
detection of individual gravitational quanta), as it is stated in the
manuscript.

5. An experimental test is suggested to check if asymmetry of the electron
angular distribution in the $\beta $ decay of light nuclei exists or not. It is not
feasible so far to provide detailed description of the experimentally observed
angular distribution in the decay of light nuclei within the frame of the 
considered approach. \emph{It is stated in the paper that there will be no asymmetry
of the distribution, and there is no experimental check of the contrary. As
for the claim that the observed angular distribution in the $\beta $-decay of heavy
nuclei "is described very well by current theories", one should note that
these theories are of matching-to-observed-values character. The values of
the Weinberg's angle determined from neutrino experiments resemble very much
the situation with the "ultraviolet catastrophe", because the numerical values
of the Weinberg's angle are simply extrapolated over the whole $\beta $-decay range.
The asymmetry of the angular distribution is experimentally proved for heavy
nuclei only}. It seems that it is expedient to carry out an experiment on light
nuclei. And if there is no asymmetry of the angular distribution of electrons
then physics will get rid of such heavy burden as the parity non-conservation
in the weak interaction.

6. All these key proposals are put forward in the article. As for an
experimental confirmation of the proposals being developed, it will lead to
the following. On one hand, it will put the gravitational interaction into
one row with the other fundamental interactions. On the other hand, it will
allow to formulate a concept of quantum generators of the gravitational
emission to be sources of high-energy states of matter (in case when the
plasma composition allows fusion reactions between plasma components) with
all the ensuing consequences (such quantum generator can also be considered
the improved accelerator of elementary particles both in terms of obtainable
energy level, and density of elementary particles beam, which is significant).
A particular case of such states, viz. the hydrostatic equilibrium states of
the dense high-temperature plasma (with absence of fusion reactions), is
discussed in the article.

Thus, the applied approach puts the gravitational interaction into one row
with the other fundamental interactions and eliminates the necessity to invent
sophisticated tests on Quantum Gravity (those are, as a rule, either virtually
non-performable, or giving no outcome). The experimental tests on Quantum
Gravity are to be performed by means of ordinary methods of elementary
particles spectroscopy, because there exist the specific quantum states of
elementary particles in their proper strong gravitational field, rather than
the general quantum states of some abstract masses. An example of such test,
being of the principle character, is the experimental check of absence of the
angular distribution asymmetry in the $\beta $-decay of light nuclei.

In case the approach studied in the manuscript is correct the expediency of
a huge number of researches towards the gravitational waves detection will
turn to be under great doubts, as well as the expediency of designing the
thermonuclear synthesis plants in the manner they are currently being designed
and built. This is the very reason explaining the negative attitude towards
the manuscript and the article in case it is published. Consequently the author
has been pushed to let the physics community get acquainted with this
manuscript by means of the Internet.

\end{document}